\documentclass[magzine]{IEEEtran}
\usepackage{amsmath,amsfonts}
\usepackage{algorithmic}
\usepackage{array}
\usepackage[caption=false,font=normalsize,labelfont=sf,textfont=sf]{subfig}
\usepackage{textcomp}
\usepackage{stfloats}
\usepackage{url}
\usepackage{verbatim}
\usepackage{graphicx}
\usepackage{xcolor}

\usepackage{setspace}
\def\BibTeX{{\rm B\kern-.05em{\sc i\kern-.025em b}\kern-.08em
    T\kern-.1667em\lower.7ex\hbox{E}\kern-.125emX}}
\usepackage{balance}
\begin{document}
\title{
Wireless AI Evolution: From Statistical Learners \\
to Electromagnetic-Guided Foundation Models
}
\author{Jian Xiao, Ji Wang, Kunrui Cao, Xingwang Li, Zhao Chen, and Chau Yuen,~\IEEEmembership{Fellow,~IEEE}
\thanks{
\emph{Corresponding author: Ji Wang.}

Jian Xiao and Ji Wang are with the Department of Electronics and Information Engineering, College of Physical Science and Technology, Central China Normal University, Wuhan 430079, China (e-mail: jianx@mails.ccnu.edu.cn; jiwang@ccnu.edu.cn). 

Kunrui Cao is with the Information Support Force Engineering University and also with the School of Information and Communications, National University of Defense Technology, Wuhan, 430035, China (e-mail: krcao@nudt.edu.cn).
	
Xingwang Li is with the School of Physics and Electronic Information Engineering, Henan Polytechnic University, Jiaozuo 454003, China (e-mail:lixingwang@hpu.edu.cn).

Zhao Chen is with the Beijing National Research Center for Information Science and Technology, Tsinghua University, Beijing 100084, China (e-mail:
zhao\_chen@tsinghua.edu.cn).

Chau Yuen is with the School of Electrical and Electronics Engineering, Nanyang Technological University, Singapore 639798 (e-mail: chau.yuen@ntu.edu.sg).
}}
%\markboth{Journal of \LaTeX\ Class Files,~Vol.~18, No.~9, September~2020}%
%{How to Use the IEEEtran \LaTeX \ Templates}
\maketitle

\begin{abstract}

While initial applications of artificial intelligence (AI) in wireless communications over the past decade have demonstrated considerable potential using specialized models for targeted communication tasks, the revolutionary demands of sixth-generation (6G) networks are propelling a necessary evolution towards AI-native wireless networks. In particular, the arrival of large AI models (LAMs) paves the way for the next phase of Wireless AI, where pre-training on universal electromagnetic (EM) principles equips wireless foundation models (WFMs) with the essential adaptability for a multitude of demanding 6G applications. However, existing LAMs face critical limitations, including pre-training strategies disconnected from EM-compliant constraints, a lack of structural adherence to wave propagation physics, and the inaccessibility of massive labeled datasets for comprehensive training. To address these challenges, this article presents an electromagnetic information theory-guided self-supervised pre-training (EIT-SPT) framework designed to systematically inject EM physics into WFMs. The EIT-SPT framework aims to infuse WFMs with intrinsic EM knowledge, thereby enhancing their physical consistency, generalization capabilities across varied EM landscapes, and overall data efficiency. Building upon the proposed EIT-SPT framework, this article first elaborates on potential applications of WFMs in 6G scenarios, then validates the efficacy of the proposed framework through illustrative case studies, and finally summarizes critical open research challenges and future directions for WFMs.

\end{abstract}

\section{Introduction}

\IEEEPARstart{T}{he} evolution toward sixth-generation (6G) wireless networks signifies a revolutionary leap forward in communication technology, far transcending the incremental upgrades of past generations \cite{10041914}. Artificial intelligence (AI) is universally recognized as a foundational technology to realize these objectives. 
Departing from traditional model-driven optimization relying on rigid mathematical frameworks, 6G necessitates a paradigm shift toward data-driven and hybrid model-data approaches. 
%Through advanced AI technologies, 6G networks will acquire self-awareness and dynamic reconfiguration capabilities.
%, enabling spectrum optimization, autonomous sensing, and other intelligent functionalities. 

\subsection{Background}
\subsubsection{From Specialized Wireless AI to Wireless {Foundation} Model}
\begin{figure*}[t]
	%		\centering
	%		\setlength{\belowcaptionskip}{-1.2cm}
	\centerline{\includegraphics[width=6.4in]{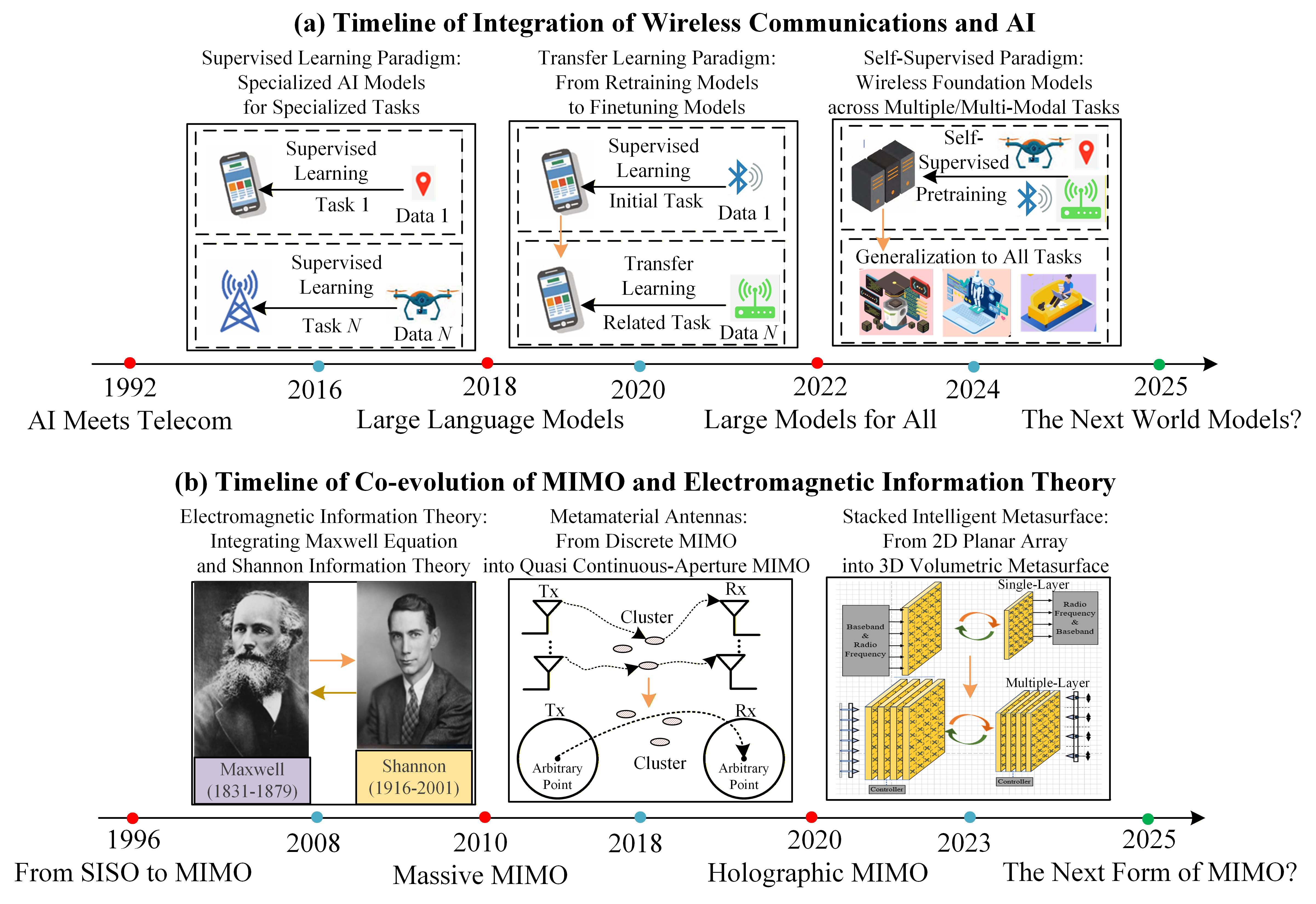}}
	%	\captionsetup{font={small},justification=raggedright,singlelinecheck=off}
	 \caption{Development of wireless AI and electromagnetic information theory.}
	\label{AI}
\end{figure*}
As depicted in Fig.~\ref{AI}(a), while the research into integrating neural networks with wireless communications began as early as 1992 \cite{153366}, the past decade has witnessed the vigorous development of Wireless AI, significantly propelled by advancements in deep learning. This initial wave was characterized by the successful deployment of specialized AI models for targeted wireless tasks. This approach inherently suffered from limited generalization and substantial customization overhead. Furthermore, challenges in tackling intricate and multi-objective problems proved to be a considerable hurdle for these specialized architectures.
The emergence and remarkable success of large AI models (LAMs), also referred to as foundation models, have paved the way for a transformative shift from task-specific learners to universal intelligent agents. These large pre-trained models demonstrate impressive generalization capabilities, which has ignited the vision for the next phase of Wireless AI: the strategic utilization of pre-trained wireless foundation models (WFMs) as a versatile and adaptable base across a multitude of diverse wireless tasks \cite{10599304}. This paradigm aims to overcome the limitations inherent in the specialized Wireless AI framework, offering improved generalization and reduced customization. {The evolution of wireless AI points towards the realization of a world model for wireless networks, an intelligent agent capable of learning an internal and predictive simulation of the physical electromagnetic (EM) environment.}

\subsubsection{From Shannon Information Theory to Electromagnetic Information Theory}

The relentless drive towards 6G networks necessitates a profound evolution in the underlying communication theories from the classic Shannon information theory that relies on abstract and statistical channel models. Fig.~\ref{AI}(b) illustrates the symbiotic evolution of multiple-input multiple-output (MIMO) technology alongside the expanding understanding of electromagnetic information theory (EIT) \cite{gruber2008new}. This co-evolutionary path highlights pivotal milestones, from MIMO inception in 1996 to the sophisticated metasurface communications of the present, necessitating a more fundamental grounding of EM wave behavior in complex propagation environments. {In particular, the advent of metamaterial antennas marks a transition from discrete to quasi-continuous apertures, necessitating EIT to characterize continuous spatial degrees of freedom \cite{10232975}. Furthermore, the evolution towards stacked intelligent metasurfaces (SIM) shifts the paradigm from planar interaction to volumetric wave processing \cite{10158690}}. {{Recognizing EIT as an evolving paradigm \cite{10417101}, this article adopts a holistic perspective that transcends statistical channel modeling by integrating Maxwell’s and Shannon’s theories. EIT incorporates EM properties to enforce physical consistency in WFMs, e.g., near-field spherical wave propagation, spatial non-stationarity across large apertures and antenna coupling of the dense array, aiming to provide a more accurate characterization of wireless link performance.}}

%The classic information theory typically relies on simplified and statistic channel models that abstract away crucial EM physics. The evolution from early MIMO systems to more advanced forms necessitates a more fundamental grounding of EM wave behavior in complex propagation environments. EIT addresses this need by integrating Maxwell's equations with information-theoretic concepts, aiming to provide a more accurate characterization of channel capacity, spatial degrees of freedom, and the impact of EM properties on wireless link performance. 

%It explicitly considers factors like near-field spherical wave propagation, spatial non-stationarity across large apertures, antenna coupling of the dense array, as well as the physics of wave interaction with materials and structures. 

\subsection {Challenges}

While the advent of LAMs offers a promising path for the evolution of Wireless AI, their direct application is hindered by the unique physics of wireless channels. Existing physics-agnostic models from natural language processing (NLP) or computer vision (CV) fail to suffice, as generic LAMs operate without structural adherence to the EM principles foundational to wireless communications. The specific challenges include:

\hangafter=1
\setlength{\hangindent}{2em}
$\bullet$ \textbf{Electromagnetic Knowledge Disconnect:} Generic LAMs lack structural adherence to EM physics, e.g., Maxwell's equations, wave propagation physics, antenna theory, or specific channel behaviors. Applying them directly can lead to physically implausible or highly suboptimal solutions for tasks requiring EM awareness.   

\setlength{\hangindent}{2em}
$\bullet$ \textbf{Physical Consistency Violation:} Physics-agnostic AI models risk violating fundamental physical laws or practical hardware constraints inherent in EM systems. This can lead to inefficient, physically unrealizable, or unreliable designs, fundamentally undermining trust and practicality in AI-driven wireless systems.

\setlength{\hangindent}{2em}
$\bullet$ \textbf{Data and Computational Inefficiency:} Training effective WFMs demands massive and diverse datasets that capture real-world EM propagation. Without the guidance of physical priors, the brute-force learning approach of LAMs necessitates even larger model capacities and more extensive datasets to implicitly capture these underlying principles, leading to unsustainable computational costs.

\subsection {Motivations and Contributions}
%Overcoming the above fundamental challenges posed by physics-agnostic LAMs in wireless communications requires a paradigm shift, i.e., the integration of underlying EM physics into the design and training lifecycle of LAMs. Simply scaling up existing models or datasets without considering the unique nature of the physical layer is unlikely to yield robust, efficient, and trustworthy WFMs. EIT provides the necessary theoretical underpinning for this integration which offers a principled way for modeling realistic EM wave propagation and interaction with the environment, characterizing the behavior of advanced MIMO systems, as well as the physical consistency with the fundamental limits considering EM constraints.
%This physics-informed approach is essential for building WFMs that are accurate, generalizable across diverse physical environments, and efficient in their operation.

To overcome the challenges posed by physics-agnostic LAMs, a fundamental evolution of wireless AI is required to embed EM physics directly into the design and training lifecycle of LAMs. Relying solely on scaling up existing LAMs fails to capture the unique physics of wireless channels. EIT bridges this gap by providing a principled theoretical basis. It provides a principled framework for modeling realistic wave propagation and characterizing advanced MIMO systems, while ensuring adherence to fundamental EM limits. This physics-grounded strategy is essential for building efficient and generalizable WFMs.

Against the above background, in this article, we develop EIT-guided WFMs that are not merely statistical learners but possess an implicit encoding of EM principles. Specifically, we propose a tri-level EIT-guided self-supervised pre-training (SPT) framework to systematically embed EM physics into WFMs. This EIT-SPT framework injects physical information through three synergistic layers: 1) an EM-compliant data genesis layer utilizing EIT to generate datasets reflecting realistic EM phenomena; 2) a physics-informed architecture layer employing model structures capable of adapting to EM field continuity and spatial dependencies; and 3) a first-principle induced pre-training layer designing SPT tasks rooted in EIT principles to instill physical causality into feature learning. We demonstrate the practical efficacy and benefits of the proposed EIT-SPT framework through compelling case studies in key 6G frontier scenarios, i.e., holographic channel estimation and high-precision wireless positioning. These case studies validate that EIT-SPT enables WFMs to achieve significant performance gains and remarkable data efficiency. Moreover, we identify and discusses critical open research issues and future directions, paving the way for continued innovation in Wireless AI.

\section{Framework of EIT-SPT Enabled Wireless Foundation Models}
To address the fundamental challenges of physics-agnosticism and data inefficiency outlined previously, this section presents the EIT-SPT framework, which is engineered to systematically embed the first principles of EM physics into the entire lifecycle of WFMs. 

\subsection{EIT for Wireless Communications}
\begin{figure}[t]
	%		\centering
	%		\setlength{\belowcaptionskip}{-1.2cm}
	\centerline{\includegraphics[width=3in]{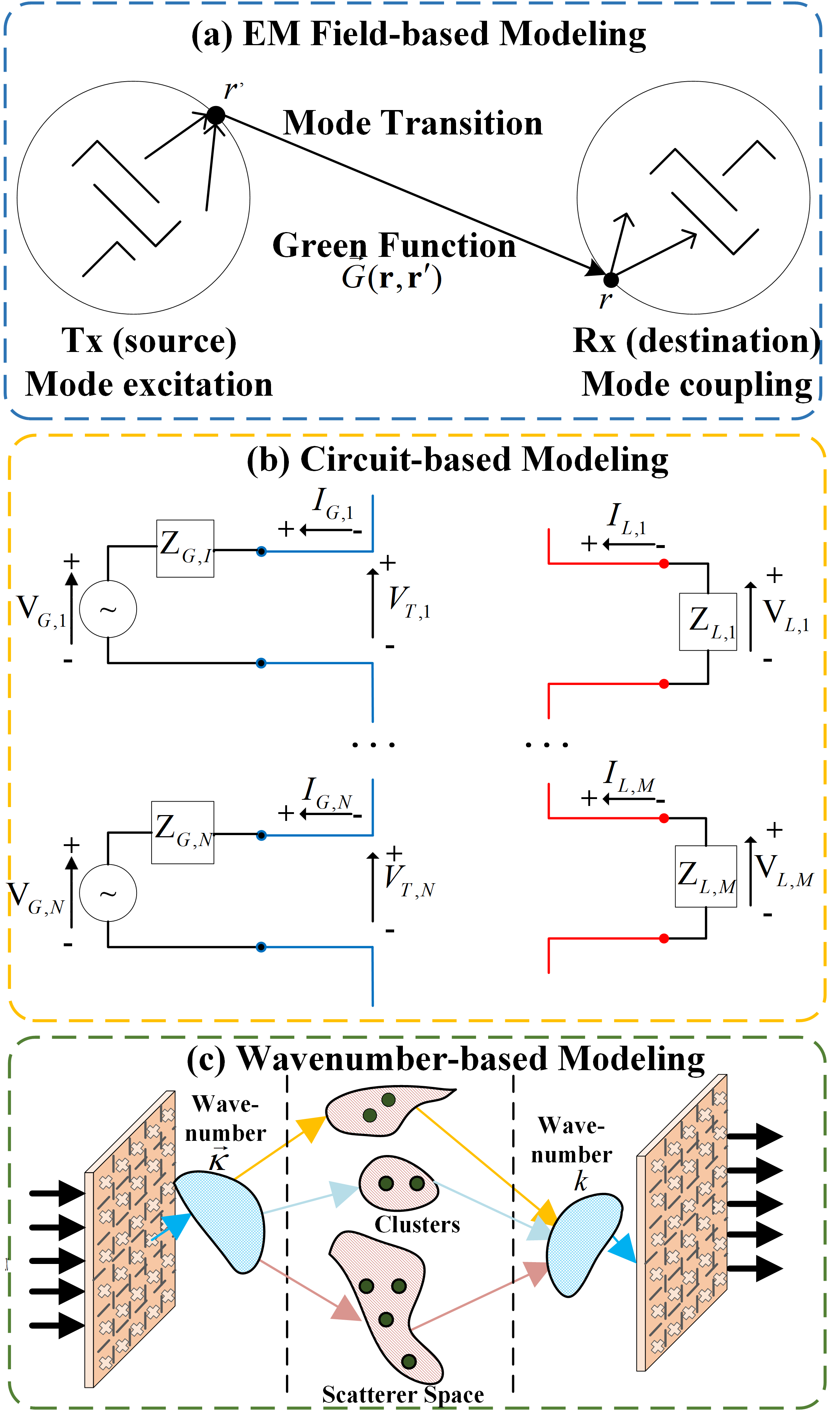}}
	%	\captionsetup{font={small},justification=raggedright,singlelinecheck=off}
	\caption{Electromagnetic information theory: (a) Field-based, (b) Circuit-based, and (c) Wavenumber-domain methods.}
	\label{EM}
\end{figure}
\subsubsection{Modeling Methods of EIT}
Fig.~\ref{EM} illustrates three complementary modeling approaches in EIT to accurately capture the EM phenomena. \textbf{(a) Field-Based Modeling:} The most fundamental approach directly models EM wave generation, propagation, and reception using Maxwell's equations, often solved with tensor Green's functions to characterize fields and environmental effects \cite{10417101}. It offers the highest fidelity for arbitrarily complex antennas and environments, naturally accounting for all wave phenomena. \textbf{(b) {Circuit-Based Modeling}}: This method abstracts EM interactions, particularly at antenna ports and within arrays, into equivalent multiport network models \cite{ivrlavc2010toward}, e.g., Z-parameters in Fig.~\ref{EM}(b). It excels at analyzing mutual coupling and impedance matching, interfacing efficiently with transceiver circuits. \textbf{(c) Wavenumber-Domain Modeling:} This approach represents fields and interactions in the spatial frequency domain via Fourier plane-wave expansion \cite{10819473}, which is well-suited for analyzing scattering and propagation in spatially extended or random media.

\subsubsection{Application Principle for EIT}
The choice among these EIT modeling methods hinges on the specific problem, desired accuracy, computational resources, and dominant EM phenomena. Field-based modeling is preferred for utmost fidelity with complex structures or near-field effects. Circuit-based modeling suits array analysis, mutual coupling, and transceiver integration. Wavenumber-domain modeling is optimal for random scattering, far-field analysis, and large-scale statistical channel characterization. In general, a hybrid approach combining elements from these methods might be the most effective way to tackle complex EIT problems, using each method where its strengths are most applicable.

\begin{figure}[t]
	%		\centering
	%		\setlength{\belowcaptionskip}{-1.2cm}
	\centerline{\includegraphics[width=3.0in]{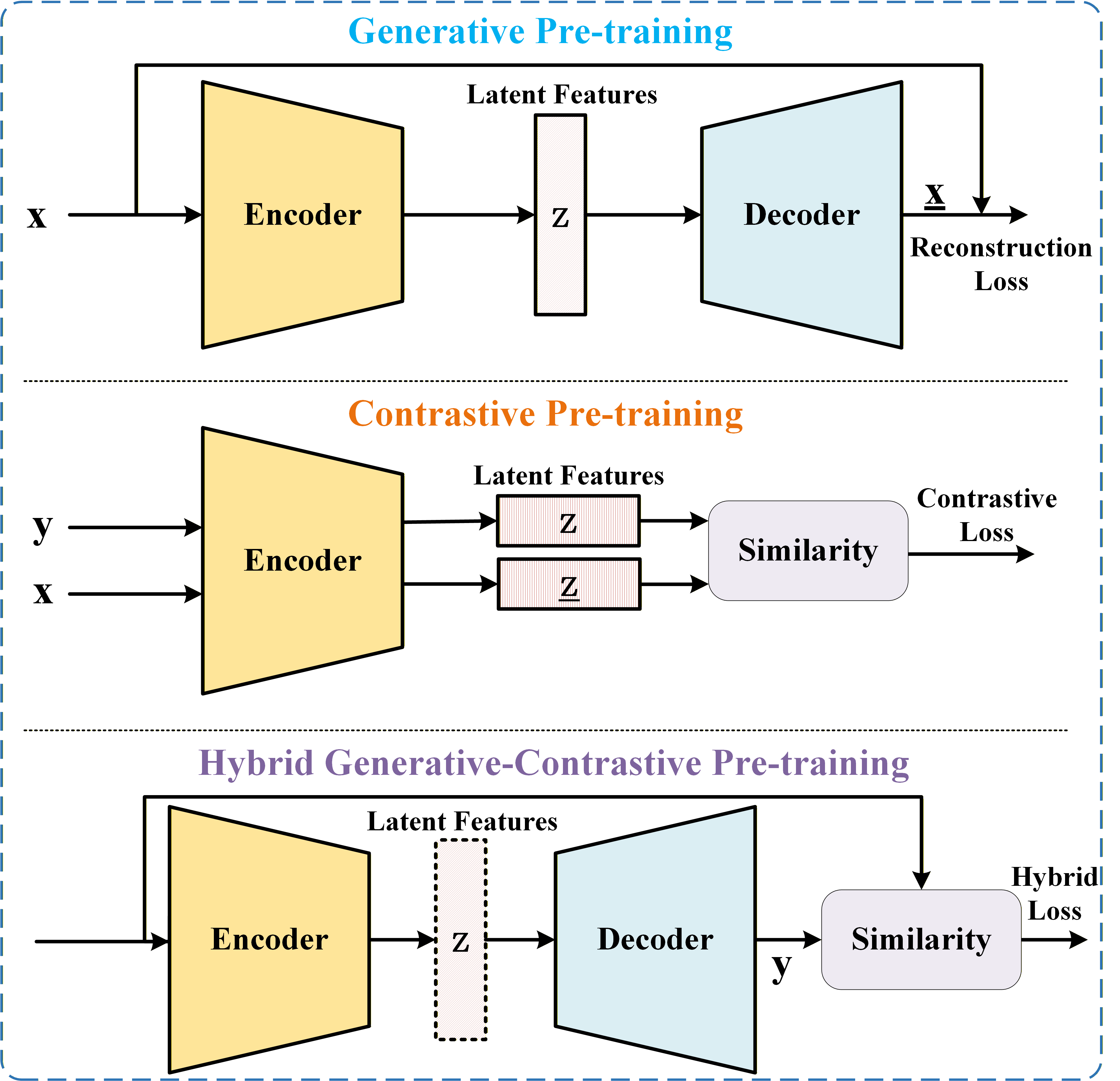}}
	%	\captionsetup{font={small},justification=raggedright,singlelinecheck=off}
	\caption{Self-supervised pre-training methods for WFMs. {Generative pre-training aims to reconstruct the original input from a latent representation. Conversely, contrastive methods learn a feature space organized by similarity, by pulling similar samples closer while pushing dissimilar ones apart. The hybrid generative-contrastive pre-training framework simultaneously integrates the reconstruction and contrastive tasks.}}
	\label{SPT}
\end{figure}

\subsection{SPT for Wireless Communications} 
\subsubsection{Learning Strategies of SPT} 
While pre-training is a foundational paradigm for LAMs, its direct application in the wireless domain is hindered by a significant challenge: the scarcity and high acquisition cost of large-scale, high-quality labeled data. This data bottleneck causes traditional supervised models to suffer from poor generalization and overfitting, limiting their reliability in dynamic wireless environments. SPT offers a powerful solution by cleverly leveraging massive unlabeled datasets, which creates its own supervisory signals from the inherent structure of the data, thereby circumventing the need for manual labels and enabling models to learn robust and generalizable feature representations. {Specifically, in the proposed EIT-SPT framework, the inherent structure refers to the governing physical dependencies within the data, e.g., spatial correlations dictated by Green’s functions and sparsity patterns in the wavenumber domain. The goal of SPT is to map these explicit physical structures into a structured latent space where geometric proximity reflects physical similarity.}
Fig.~\ref{SPT} presents the generative, contrastive and hybrid generative and contrastive pre-training frameworks \cite{9462394}. 

%Generative approaches aim to reconstruct the original input from a latent representation, making them highly effective for tasks requiring data structure comprehension or synthesis, like channel estimation and completion. Contrastive methods, conversely, learn a feature space organized by similarity, by pulling similar samples closer while pushing dissimilar ones apart. This makes them particularly well-suited for discriminative downstream tasks. {By simultaneously minimizing a reconstruction loss and a contrastive loss, the generative-contrastive frameworks models can achieve a holistic understanding of the EM environment. This synergy allows the WFM to capture the detailed wave propagation physics while maintaining a structured latent space that is robust to noise and environmental variations.}

\subsubsection{Application Principle of SPT for Wireless Communications}
The selection of an appropriate SPT method for a specific wireless communication task hinges on aligning the pre-training objective with the requirements of the intended downstream application. Generative approaches are generally preferred for tasks focused on data reconstruction, completion, denoising or generation, e.g., channel estimation, signal denoising or synthetic data generation, as their learned latent variables encapsulate rich generative information. Conversely, contrastive approaches tend to perform better when the downstream task is inherently about classification, identification, or learning highly discriminative features. For example, in modulation recognition, interference type identification, or specific radio frequency (RF) fingerprinting tasks, the class invariance and feature separability are paramount. {The hybrid generative-contrastive frameworks are ideal for complex and multifaceted scenarios where the model must simultaneously maintain structural physical fidelity and capture robust high-level semantics, e.g., preserving signal integrity and distinguishing environmental contexts.}
\subsection {EIT-SPT: Tri-Level Hierarchical Framework}
\begin{figure*}[t]
	%		\centering
	%		\setlength{\belowcaptionskip}{-1.2cm}
	\centerline{\includegraphics[width=6.2in]{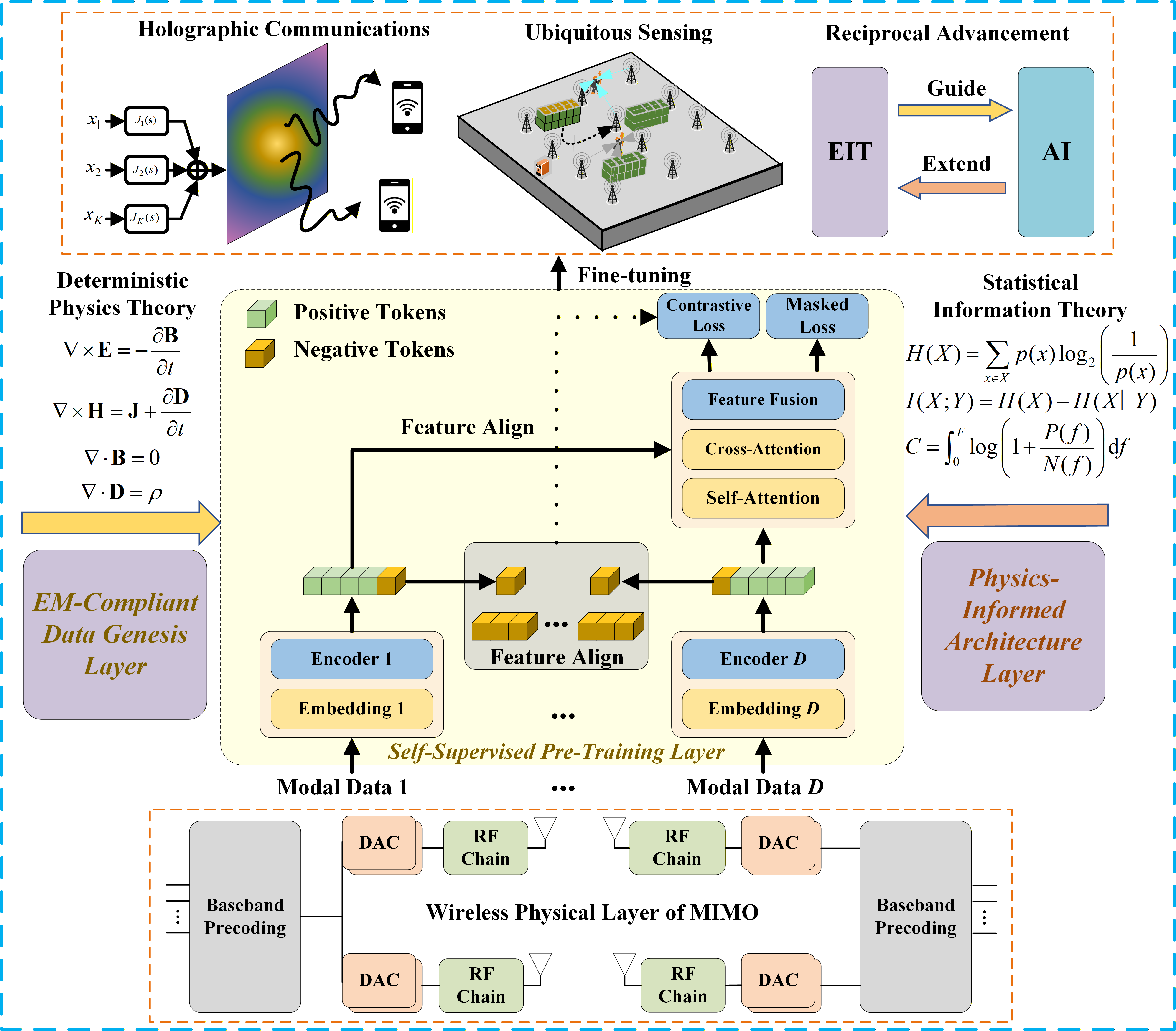}}
	%	\captionsetup{font={small},justification=raggedright,singlelinecheck=off}
	\caption{The proposed EIT-SPT framework for WFMs. {It systematically injects physical laws into the WFM lifecycle through three synergistic layers: 1) The EM-compliant data genesis layer ensures the input follows physical laws; 2) The physics-informed architecture layer provides the necessary structural inductive bias for continuous fields; and 3) The self-supervised pre-training layer employs physics-rooted tasks to enforce causal learning.}}
	\label{EIT}
\end{figure*}

As illustrated in Fig.~\ref{EIT}, the proposed EIT-SPT framework that embeds EIT into the lifecycle of WFMs, aims to create models that are not only powerful learners but also inherently understand the EM environment they operate in. {While EIT provides the structural inductive bias, statistical information theory complements this by governing the optimization process to ensure efficient information flow and compression within the physical architecture.}

\subsubsection {EM-Compliant Data Genesis Layer}
    The efficacy of any LAM heavily relies on vast and diverse datasets. However, the wireless domain suffers from a scarcity of universal, physics-aware data. Generic datasets fail to capture the nuances of EM wave propagation. Our framework addresses this by championing the use of EIT-based simulation methods to generate large-scale, synthetic datasets. These simulations, grounded in Maxwell's equations and potentially employing techniques like tensor Green's functions to link source currents to received fields, can produce high-fidelity channel data. Such data inherently respects fundamental EM laws, capturing complex phenomena like diffraction, scattering, near-field wave curvature, spatial non-stationarity across large arrays, and antenna coupling. Pretraining WFMs on this EIT-generated data endows them with a foundational grasp of EM physics.

\subsubsection {Physics-Informed Architecture Layer}
    Directly adopting LAM architectures from fields like NLP or CV is often suboptimal for wireless tasks due to their disregard for underlying physical principles. Standard wireless models often view signals and channels discretely. In contrast, EIT highlights the continuous nature of EM fields. Our framework advocates for designing LAM architectures that are inherently physics-informed. This involves creating models that can natively handle or approximate spatially continuous fields, crucial for continuous-aperture antenna arrays. An EIT-informed architecture might explicitly model spatial dependencies or utilize representations suited for continuous EM fields, thus bridging the gap between discrete computation and continuous physics and addressing the lack of EM knowledge and physical consistency. {The physics-informed architecture layer functions as a universal backbone for extracting general EM features, upon which tailored task-specific heads are designed to address distinct downstream applications.}
    
    \subsubsection {Self-Supervised Pre-Training Layer} This layer achieves this by formulating SPT tasks that are rooted in EIT principles, rather than relying on generic learning objectives. For instance, we can compel the model to perform wave equation-constrained masked autoencoding on EM field data, and then reconstruct wavenumber channel representations consistent with wave physics. By requiring WFMs to learn features that inherently align with fundamental EM laws and interactions, this layer enhances its ability to generalize across diverse and previously unseen EM environments and contributes to more efficient learning by embedding strong physical priors, thereby reducing the need to learn these principles purely from data.
    
    {In summary, the proposed EIT-SPT framework bridges the gap between statistical learning and physical reality, which is established through the statistical approximation of physical operators. By training on data strictly constrained by the Green’s function manifold in data layer and minimizing reconstruction objectives that enforce spatial causality in pre-training layer, the WFM is compelled to internalize the governing wave equations as latent feature representations by physics-informed network architecture. This allows the statistical model to approximate the deterministic propagation rules of the physical environment, ensuring consistency with EM laws. It is worth distinguishing the proposed EIT-SPT framework from the conventional physics-informed neural networks (PINNs). While PINNs typically function as neural solvers that minimize residuals of partial differential equations for specific instances, the EIT-SPT framework operates as a foundation model learner that internalizes physical laws into universal latent representations via EIT-compliant pre-training.}

\subsection{Fundamental Applications of WFMs in Wireless AI}
\subsubsection{WFMs for Holographic Communications}

Holographic MIMO communications, operating in the near-field with continuous or near-continuous apertures, present significant challenges that physics-agnostic AI models struggle with. The EIT-SPT enabled WFMs offer targeted solutions for specific EM physics, which can perform electromagnetically consistent near-field channel estimation, accurately modeling spherical wave effects and spatial non-stationarity phenomena. Furthermore, WFMs can achieve the design of physics-aware beamforming and focusing, generating physically realizable and highly directive patterns consistent with array characteristics. 

\subsubsection{WFMs for Ubiquitous Sensing}

Integrated Sensing and Communications is a cornerstone of 6G, requiring the network to accurately perceive its surroundings by interpreting how communication signals interact with the environment. WFMs leverages its fundamental physics knowledge for enhanced environmental perception. Pretrained with an understanding of EM scattering, reflection, and diffraction principles, it facilitates high-fidelity interpretation of EM signatures, robustly decoding subtle signal variations like micro-Doppler shifts caused by interactions with objects and humans. Building upon this, WFMs can establish an EM-compliant understanding of the radio scene and reliably detect anomalies by identifying deviations from a physically-grounded baseline.

\subsubsection{WFMs for EIT Evolution}

While EIT provides the indispensable physical grounding for constructing WFMs, this relationship is not unidirectional but forms a virtuous cycle where WFMs, augmented with potent AI capabilities, can in turn accelerate the evolution of EIT itself. Specifically, WFMs can act as highly efficient EM solver emulators, learning from simulation data to rapidly predict complex EM field distributions and channel characteristics, thereby accelerating traditionally time-consuming numerical computations. Moreover, {WFMs can facilitate the characterization and discovery of complex EM behaviors by serving as efficient surrogate models for high-dimensional parameter exploration. This capability allows for the rapid interrogation of the entire EIT solution manifold, enhancing the solving of ill-posed EM inverse problems from sparse measurements.}

%The WFMs can also enhance the solving of ill-posed EM inverse problems by learning from extensive data pairings and incorporating EIT-derived physical constraints to robustly infer parameters from sparse measurements.

\section{Case Studies in Communications \& Sensing}

\begin{figure*}[t]
	%		\centering
	%		\setlength{\belowcaptionskip}{-1.2cm}
	\centerline{\includegraphics[width=7in]{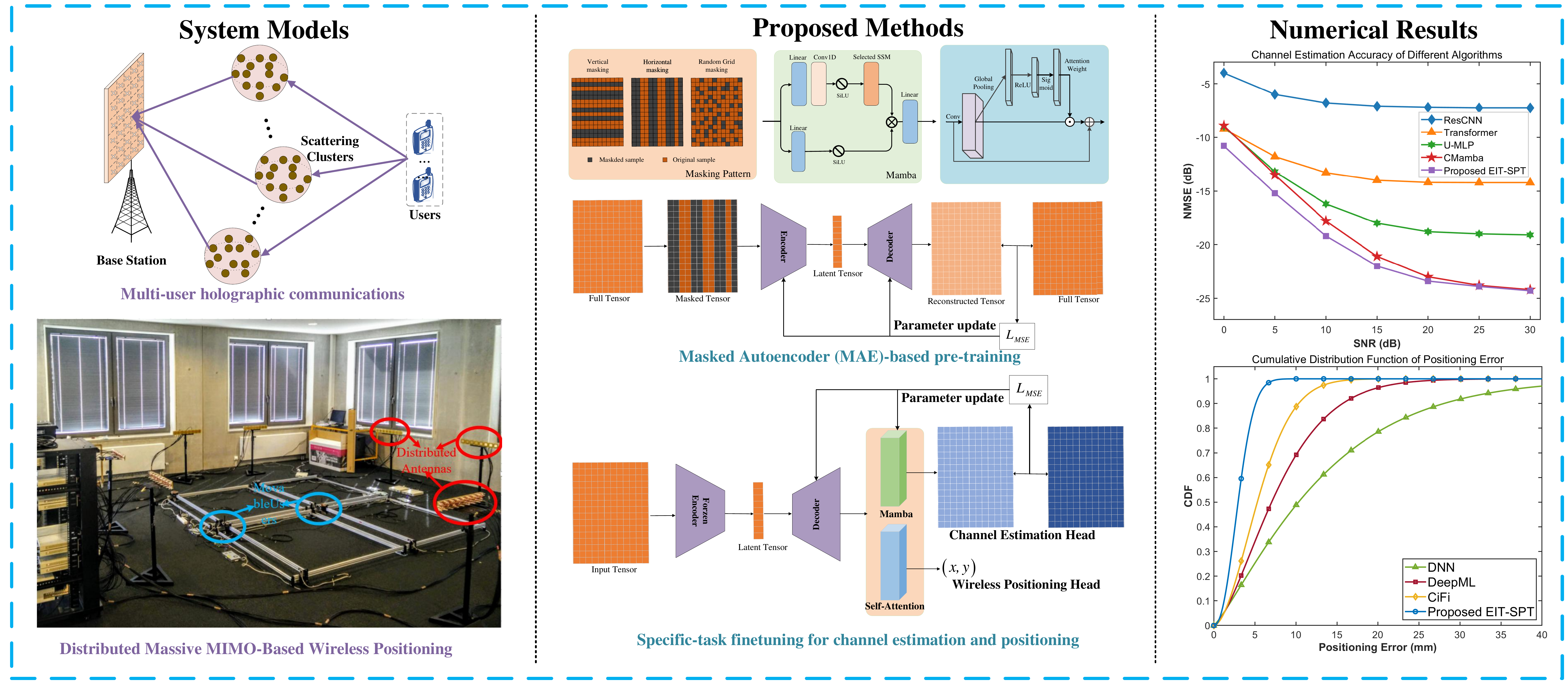}}
	%	\captionsetup{font={small},justification=raggedright,singlelinecheck=off}
	\caption{Case studies for channel estimation and positioning of the proposed EIT-SPT framework. {For the channel estimation case, we consider a multi-user HMIMO system, where a base station comprises densely packed uniform planar array antennas with sub-wavelength spacing. For the wireless positioning case, the CSI dataset is  measured via a real-world distributed Massive MIMO testbed, where the total grid of positioning area spans 1.25 m by 1.25 m \cite{9129126}.}}
	\label{Case}
\end{figure*}

To concretize the potential benefits of the proposed EIT-SPT framework, this section provides typical case studies for EIT-SPT enabled WFMs. The considered system models, proposed methods and numerical results are depicted in Fig.~\ref{Case}.

\subsection{EIT-SPT for Holographic Channel Estimation}
The massive antennas and spatially near-continuous apertures in holographic MIMO (HMIMO) systems cause traditional channel estimation schemes to require larger pilot overhead. In particular, for AI-based channel estimation approaches, acquiring sufficient labeled data for training deep learning models is difficult and costly. In this work, we leverage the proposed EIT-SPT framework to achieve the accurate holographic channel estimation with limited labeled training samples. {Specifically, we construct an EM-compliant data representation rooted in EIT, where the HMIMO channel is characterized in the wavenumber domain. This approach strictly adheres to EIT principles by rigorously capturing the physical wave propagation and continuous field distributions inherent to holographic aperture arrays.} {Due to the finite aperture effect and the Fourier duality inherent in HMIMO system, the spectral leakage across the wavenumber domain causes the energy of a single propagation path to spread globally rather than being localized to a single point, creating intrinsic long-range spectral correlations.} {To specifically address the spatially continuous nature of holographic fields, the selective state space model (SSM)-based Mamba module is employed to construct a physics-informed network architecture \cite{gu2023mamba}. By modeling the spatial dimension as a continuous dynamic process, the Mamba captures the smooth phase variations and long-range diffractive correlations inherent to the holographic aperture, effectively bridging the gap between discrete digital processing and continuous physical reality.} Furthermore, to address labeled data scarcity, we construct a masking-based pre-training task, allowing the model to learn robust feature representations from unlabeled wavenumber channel data by reconstructing the masked channels with different masking pattern to the full channels. 
 
{The numerical results of channel estimation in Fig.~\ref{Case} show that the proposed EIT-SPT framework achieves significantly lower normalized mean squared error (NMSE) compared to the existing schemes. In particular, the proposed EIT-SPT only utilizes 3000 training samples in the network training stage, while the existing models, e.g., ResCNN, Transformer and U-MLP, require 20000 training samples \cite{11090043}.}
This is because the proposed model leverages its broad knowledge of EIT-consistent channel characteristics learned during SPT. In particular, the proposed EIT-SPT framework effectively achieves superior accuracy with limited labeled samples, which demonstrates the power of the EIT-SPT framework in developing powerful and data-efficient WFMs for specialized wireless tasks.

\subsection{EIT-SPT for Wireless Positioning}
{In this case study, the measured channel state information (CSI) dataset provides ultra-dense spatial channel sampling measurements with a 5mm step size. This sub-wavelength granularity effectively acts as a high-resolution discretization of the continuous EM field.} We utilize the proposed EIT-SPT framework to understand the information within these EM signatures, relating path delays and amplitudes to the physical environment and user location, where the effective masking patterns in the antenna domain of CSI and the attention modules are applied. 
The model learns latent representations by reconstructing the original CSI from these masked inputs using unlabeled CSI samples, effectively learning the underlying structure shaped by EM physics. {In particular, we incorporate the channel attention (CA) mechanism to the backbone of the proposed EIT-SPT for wireless positioning \cite{10897531}. According to EIT, the CSI is not a uniform signal but consists of sparse multipath components amidst noise. Standard convolutional networks treat all extracted features equally, which is physically inefficient. By integrating CA, we introduce an adaptive mode selection capability. The CA module dynamically re-weights the feature maps, effectively distinguishing and amplifying the features corresponding to physically valid propagation paths, i.e., energy-concentrated subspaces, while suppressing those related to environmental noise or measurement artifacts.} 
{The position results in Fig.~\ref{Case} present the cumulative distribution function (CDF) of positioning error of different positional network models, where the proposed EIT-SPT framework shows superior positioning accuracy compared to existing models \cite{10678770}.}

\section{Open Issues and Future directions}
The development and deployment of WFMs represent a significant undertaking, presenting numerous open research questions and challenges that need to be addressed to fully realize the Wireless AI evolution.

\subsection{Sustainable WFMs with Collaborative Learning}
The substantial computational demands for training and operating large WFMs present serious sustainability challenges, making it crucial to minimize their energy consumption and carbon footprint. The necessary complexity for physical accuracy needs to be balanced with operational energy efficiency. Hence, developing energy-efficient distributed learning techniques like federated collaborative learning, specifically tailored for EIT-based models, will be essential. For instance, the physical locality inherent in EM propagation could inform more efficient federated averaging or client selection strategies. 

\subsection{ Trustworthy WFMs with Adversarial Machine Learning}
Ensuring the reliability, robustness, and security of foundation models controlling critical communication infrastructure is paramount. Adversarial attacks pose a significant threat, where WFMs need to be robust against adversarial manipulations targeting either their training data, and ensure their decisions are interpretable and fair. Consequently, research efforts should focus on developing adversarial detection and defense mechanisms that leverage EIT-derived physical consistency checks, e.g., by detecting if an input forces the model to predict a physically impossible EM field. 

\subsection{ Scalable WFMs with Continual Learning}
While the EIT-SPT framework endows WFMs with foundational EM knowledge, real-world wireless environments are inherently dynamic, and network requirements constantly evolve. A critical challenge for long-term scalability and relevance is enabling EIT-SPT based WFMs to efficiently adapt to new data, tasks, or changing EM conditions post-training without catastrophic forgetting or the need for complete retraining. A key research thrust will be investigating continual learning strategies tailored for physics-informed WFMs. This involves creating methods for efficient post-training adaptation and specialization while preserving core EIT principles. 

\section{Conclusion}
This article presented the EIT-SPT framework to construct efficient WFMs, which is designed to bridge the fundamental disconnect between generic LAMs and the physical realities of wireless communications. By injecting EM laws across three synergistic layers, including data genesis, architecture design and pre-training principles, we transformed WFMs from a mere statistical learner into a physics-grounded entity capable of implicit physical encoding. The presented case studies in holographic channel estimation and wireless positioning empirically validated that this framework significantly enhances model performance and data efficiency compared to conventional baselines. The fusion of EIT with AI scalability lays the essential groundwork for realizing trustworthy, sustainable, and high-performance AI-native 6G networks.

%The quest for 6G necessitates a fundamental rethinking of how AI is integrated into wireless systems, moving beyond the limitations of task-specific models towards the universal adaptability promised by WFMs. However, this transition cannot succeed by simply importing AI paradigms from other domains. Wireless communication is intrinsically governed by the laws of electromagnetism, and ignoring this physical reality leads to models that lack robustness, generalizability, and trustworthiness. This article has strongly advocated for the integration of EIT as the cornerstone for designing the next generation of wireless foundation models. By grounding AI models in the fundamental physics described by EIT, this article paved the way for WFMs that are not only powerful statistical learners but also possess an EM-compliant representations of the medium they operate in. This leads to enhanced performance in emerging 6G scenarios and provides a more solid foundation for building sustainable and trustworthy AI-driven networks. The synergistic fusion of EIT physical rigor and AI learning power is key to unlocking the full potential of intelligent, adaptable, and performant 6G systems and truly realizing AI-native wireless networks.
\bibliographystyle{IEEEtran}
\bibliography{IEEEabrv,refs.bib}

% Generated by IEEEtran.bst, version: 1.14 (2015/08/26)
\begin{thebibliography}{10}
\providecommand{\url}[1]{#1}
\csname url@samestyle\endcsname
\providecommand{\newblock}{\relax}
\providecommand{\bibinfo}[2]{#2}
\providecommand{\BIBentrySTDinterwordspacing}{\spaceskip=0pt\relax}
\providecommand{\BIBentryALTinterwordstretchfactor}{4}
\providecommand{\BIBentryALTinterwordspacing}{\spaceskip=\fontdimen2\font plus
\BIBentryALTinterwordstretchfactor\fontdimen3\font minus
  \fontdimen4\font\relax}
\providecommand{\BIBforeignlanguage}[2]{{%
\expandafter\ifx\csname l@#1\endcsname\relax
\typeout{** WARNING: IEEEtran.bst: No hyphenation pattern has been}%
\typeout{** loaded for the language `#1'. Using the pattern for}%
\typeout{** the default language instead.}%
\else
\language=\csname l@#1\endcsname
\fi
#2}}
\providecommand{\BIBdecl}{\relax}
\BIBdecl

\bibitem{10041914}
M.~Chafii, L.~Bariah, S.~Muhaidat, and M.~Debbah, ``Twelve scientific
  challenges for {6G}: Rethinking the foundations of communications theory,''
  \emph{IEEE Commun. Surv. Tutorials}, vol.~25, no.~2, pp. 868--904,
  Secondquarter 2023.

\bibitem{153366}
B.~Aazhang, B.-P. Paris, and G.~Orsak, ``Neural networks for multiuser
  detection in code-division multiple-access communications,'' \emph{IEEE
  Trans. Commun.}, vol.~40, no.~7, pp. 1212--1222, Jul. 1992.

\bibitem{10599304}
S.~Xu, C.~Kurisummoottil~Thomas, O.~Hashash, N.~Muralidhar, W.~Saad, and
  N.~Ramakrishnan, ``Large multi-modal models {(LMMs)} as universal foundation
  models for {AI}-native wireless systems,'' \emph{IEEE Network}, vol.~38,
  no.~5, pp. 10--20, Sep. 2024.

\bibitem{gruber2008new}
F.~K. Gruber and E.~A. Marengo, ``New aspects of electromagnetic information
  theory for wireless and antenna systems,'' \emph{IEEE Trans. Antennas
  Propag.}, vol.~56, no.~11, pp. 3470--3484, Nov. 2008.

\bibitem{10232975}
T.~Gong, P.~Gavriilidis, R.~Ji, C.~Huang, G.~C. Alexandropoulos, L.~Wei,
  Z.~Zhang, M.~Debbah, H.~V. Poor, and C.~Yuen, ``Holographic {MIMO}
  communications: Theoretical foundations, enabling technologies, and future
  directions,'' \emph{IEEE Commun. Surv. Tutorials}, vol.~26, no.~1, pp.
  196--257, Firstquarter 2024.

\bibitem{10158690}
J.~An, C.~Yuen, C.~Xu, H.~Li, D.~W.~K. Ng, M.~Di~Renzo, M.~Debbah, and
  L.~Hanzo, ``Stacked intelligent metasurface-aided {MIMO} transceiver
  design,'' \emph{IEEE Wireless Commun.}, vol.~31, no.~4, pp. 123--131, Aug.
  2024.

\bibitem{10417101}
J.~Zhu, Z.~Wan, L.~Dai, M.~Debbah, and H.~V. Poor, ``Electromagnetic
  information theory: Fundamentals, modeling, applications, and open
  problems,'' \emph{IEEE Wireless Commun.}, vol.~31, no.~3, pp. 156--162, Jun.
  2024.

\bibitem{ivrlavc2010toward}
M.~T. Ivrla{\v{c}} and J.~A. Nossek, ``Toward a circuit theory of
  communication,'' \emph{IEEE Trans. Circuits Syst. I Regul. Pap.}, vol.~57,
  no.~7, pp. 1663--1683, Jun. 2010.

\bibitem{10819473}
Y.~Chen, X.~Guo, G.~Zhou, S.~Jin, D.~W.~K. Ng, and Z.~Wang, ``Unified far-field
  and near-field in holographic {MIMO}: A wavenumber-domain perspective,''
  \emph{IEEE Commun. Mag.}, vol.~63, no.~1, pp. 30--36, Jan 2025.

\bibitem{9462394}
X.~Liu \emph{et~al.}, ``Self-supervised learning: Generative or contrastive,''
  \emph{IEEE Trans. Knowl. Data Eng.}, vol.~35, no.~1, pp. 857--876, Jun. 2023.

\bibitem{9129126}
S.~D. Bast, A.~P. Guevara, and S.~Pollin, ``{CSI}-based positioning in massive
  mimo systems using convolutional neural networks,'' in \emph{Proc. IEEE 91st
  Vehicular Technology Conference (VTC2020-Spring)}, 2020.

\bibitem{gu2023mamba}
A.~Gu and T.~Dao, ``Mamba: Linear-time sequence modeling with selective state
  spaces,'' \emph{arXiv preprint arXiv:2312.00752}, 2023.

\bibitem{11090043}
\BIBentryALTinterwordspacing
J.~Wang \emph{et~al.}, ``Deep learning based wavenumber domain channel
  estimation for holographic {MIMO} communications,'' \emph{IEEE Trans. Veh.
  Technol.}, 2025. [Online]. Available: \url{doi: 10.1109/TVT.2025.3591440}
\BIBentrySTDinterwordspacing

\bibitem{10897531}
M.-H. Guo \emph{et~al.}, ``Attention mechanisms in computer vision: A survey,''
  \emph{Comput. Visual Media}, vol.~8, no.~3, pp. 331--368, Sep. 2022.

\bibitem{10678770}
J.~Wang, W.~Fang, J.~Xiao, Y.~Zheng, L.~Zheng, and F.~Liu, ``Signal-guided
  masked autoencoder for wireless positioning with limited labeled samples,''
  \emph{IEEE Trans. Veh. Technol.}, vol.~74, no.~1, pp. 1759--1764, Jan. 2025.

\end{thebibliography}
%{ % 开始局部作用域
%\setlength{\parskip}{0pt}
%\begin{IEEEbiographynophoto}{Jian Xiao}
% is a Ph.D. candidate with the Department of Electronics and Information Engineering Central China Normal University.
%\end{IEEEbiographynophoto}
%\vspace{-5mm} % 减少 5 毫米间距
%
%\begin{IEEEbiographynophoto}{Ji Wang}
% is a associate professor with the Department of Electronics and Information Engineering Central China Normal University.
%\end{IEEEbiographynophoto}
%\vspace{-5mm} % 减少 5 毫米间距
%
%\begin{IEEEbiographynophoto}{Kunrui Cao}
% is a associate professor with the School of Information and Communications, National University of Defense Technology.
%\end{IEEEbiographynophoto}
%\vspace{-5mm} % 减少 5 毫米间距
%
%\begin{IEEEbiographynophoto}{Xingwang Li}
% is a associate professor with the School of Physics and Electronic Information Engineering, Henan Polytechnic University.
% \end{IEEEbiographynophoto}
% \vspace{-5mm} % 减少 5 毫米间距
%
%\begin{IEEEbiographynophoto}{Zhao Chen}
% is a associate professor with the Beijing National Research Center for Information Science and Technology, Tsinghua University.
% \end{IEEEbiographynophoto}
% \vspace{-5mm} % 减少 5 毫米间距
%
%\begin{IEEEbiographynophoto}{Chau Yuen}
% is a associate professor with the School of Electrical and Electronics Engineering, Nanyang Technological University
%\end{IEEEbiographynophoto}}
%\vspace{-5mm} % 减少 5 毫米间距

%
%\begin{IEEEbiography}[{\includegraphics[width=1in,height=1.25in,clip,keepaspectratio]{fig1.png}}]{IEEE Publications Technology Team}
%In this paragraph you can place your educational, professional background and research and other interests.\end{IEEEbiography}

\end{document}